# CG Draconis – a particularly active dwarf nova

Jeremy Shears, Roger Pickard & Gary Poyner

The dwarf nova CG Dra was intensively monitored during 2005 as part of the Variable Star Section's Recurrent Objects Programme and seven outbursts were detected. These observations were combined with others from the BAA and AAVSO databases which confirmed that the star shows frequent outbursts with a period of around 11 days. Two types of outburst have been detected: short outbursts lasting about 4 days and long outbursts lasting about 8 days.

## Background

CG Dra is a comparatively little known dwarf nova discovered by Hoffmeister in 1965.[1] In a later paper,[2] Hoffmeister classified it as U Gem-like dwarf nova with frequent outbursts.

Albert Bruch and his team observed CG Dra photographically over eight nights and found it once in outburst.[3] Subsequently, a report by astronomers at Kyoto University resulting from a campaign in 1996, suggested an outburst frequency of less than 82 days.[4] This was based on two observed outbursts, although the observations were limited and each outburst was only caught during the declining phase. All-in-all, very few observations of the star exist in the literature, so its outburst behaviour remains poorly characterised. The General Catalogue of Variable Stars[5] lists CG Dra as having a photographic magnitude range of 15 to 17.5.

Dwarf novae are binary stars comprising a cool main sequence star or red dwarf (the secondary star) orbiting a white dwarf (primary). Matter flows from the secondary towards the white dwarf and forms an accretion disc around it. From time to time the accretion disc flips between a dimmer, cooler state to a hotter, brighter state, resulting in what we see as an outburst. A spectroscopic study by Bruch revealed two spectral components, which are attributed to the accretion disc and the secondary, whose characteristics are consistent with CG Dra being a dwarf nova.[6] However, there are a number of peculiarities in the spectrum which are difficult to explain by the standard models for dwarf novae, and Bruch could not provide a self-consistent description of the system. His radial velocity measurements indicate an orbital period of either 4h 32min or 5h 37min.

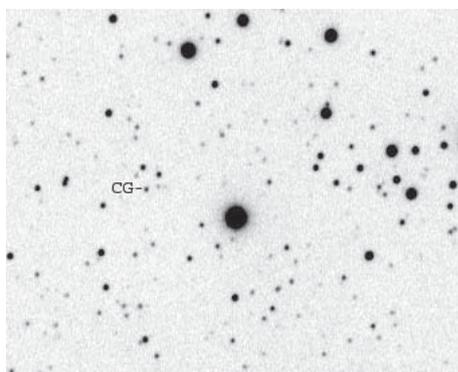

**Figure 1.** CG Dra in outburst. 2005 April 9, 00.11UT. Takahashi FS102, 0.1m refractor. 60 sec. image with unfiltered Starlight Xpress SXV-M7 CCD. Field 11.5´×9´, south at top, east to right. *(Jeremy Shears)*

## CG Dra and the Recurrent Objects Programme

CG Dra was added to the BAA Variable Star Section's Recurrent Objects Programme (ROP) in 2001.[7] The aim of the ROP is to encourage monitoring of poorly characterised dwarf novae and other stars. Charts for CG Dra are available from both the BAA VSS and the American Association of Variable Star Observers (AAVSO) web sites.[8,9]

An intensive campaign was conducted by several VSS members during 2005: David Boyd, Roger Pickard and Jeremy Shears carried out CCD surveillance, whilst Gary Poyner and Chris Jones contributed visual observations. The first outburst of the year was detected by JS on 2005 April 9 at 16.09C (Figure 1). During the period 2005 April 9 to October 22, seven outbursts were reported by the team (Table 1). This indicates that CG Dra is indeed a dwarf nova, with an outburst period which is considerably shorter than that found by the Kyoto team.[4]

Although several outbursts were detected, it is possible that others were missed due to incomplete coverage resulting from periods of poor weather over the UK. To obtain a more complete data set, covering a longer period of time, observations were requested from both the BAA VSS and the AAVSO databases. A total of 1636 observations was forthcoming, covering the period 1995 to 2005. These include visual observations, unfiltered CCD, CCD with V filter (CCD+V, approximating visual magnitudes) and CCD with B filter (CCD+B). Of these, 1153 were positive observations and these are plotted in Figure 2. It is evident that comparatively few positive observations exist prior to JD2453100 (2004 March), whereas there was better coverage towards the end of the period. This is mainly because publicity was given to CG Dra by the present authors on various variable star web sites during 2005, which resulted in additional observations being submitted by other observers.

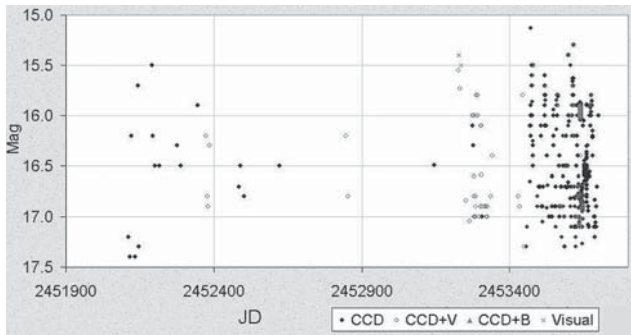

**Figure 2.** Positive AAVSO and BAA observations during the period JD 2451900 to 2453700 (2000 December to 2005 November 25). Only two further positive observations, from 1996 July 17 (16.5C) and 1997 April 6 (16.4C), exist in these databases. They have been omitted for clarity, as have all negative observations.

In an attempt to make the outbursts clearer, the visual and CCD data covering the period JD 2453456 to 2453700 (2005 March 26 to Nov 25) were combined and replotted as Figure 3. Again a multiplicity of outbursts can be seen, with an extreme outburst amplitude of mag 15.1 to 17.4. It is evident that CG Dra spends little time at quiescence. If it is assumed that anything above mag 16.5 is an outburst, then it is possible to identify 15 separate outbursts during this period (Table 2). There is a wide range of outburst intervals, with a median of outburst period of 11 days. The shortest interval between outbursts is 5 days and the longest is 35 days (however the latter could be due to outbursts having been missed due to incomplete observational coverage). Statistical analysis using a variety of methods was attempted using the *Peranso* software,[10] but in all cases failed to yield a definitive outburst period.

The approximate duration of each outburst is also listed in Table 2. In some cases it was not possible to determine the end of the outburst due to incomplete observational coverage; these outbursts have been identified as U, 'undetermined'. It appears that some outbursts have a double maximum. However, this is most likely an artefact due to combining all the observations made by different detection methods (i.e. visual, unfiltered CCD and filtered CCD), each having a different spectral response. Hence any double peaked outburst is assumed to be a single outburst. This is supported by the fact that there was no decline below 16.5 between the maxima. There is a wide range of outburst durations: the shortest outburst lasted

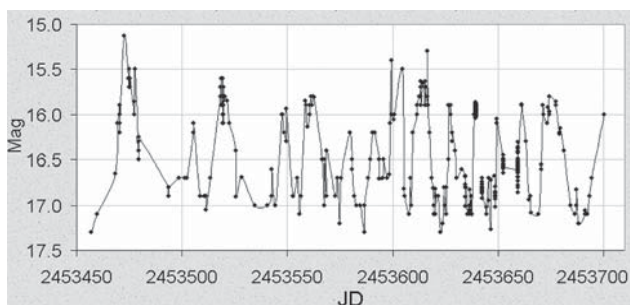

**Figure 3.** AAVSO and BAA observations during the period JD 2453456 to 2453700 (2005 March 26 to November 25). In this plot, visual and CCD observations are combined. Negative observations have been omitted for clarity.

**Table 1. Outbursts reported to the BAAVSS between 2005 April 9 and 2005 October 22**

| Date (2005) | Observer | Mag | Confirmed |
|---|---|---|---|
| Apr 09 | Shears | 16.1C | Pickard |
| May 27 | Shears | 15.8C | Poyner |
| July 06 | Pickard | 15.8V | Shears/Poyner |
| Aug 16 | Pickard | 15.7V | Poyner |
| Aug 29 | Shears | 15.8C | Pickard |
| Sep 13 | Shears | 15.9C | |
| Oct 17 | Shears | 16.0C | |

1 day and the longest 11 days, with an average of 5.6 days. Figure 4 displays a histogram of the outburst durations, which appears to show a bi-modal distribution with a peak at 4 days and another at 8 days. Whilst this is by no means definitive, it is in line with the distribution of short and long outbursts seen in some other dwarf novae.[11]

## Conclusion

The observations reported in this paper indicate that CG Draconis is a dwarf nova with an outburst period of around 11 days. The star appears to be particularly active and spends very little time at quiescence. Two types of outburst have been detected: short outbursts lasting about 4 days and long outbursts lasting about 8 days.

In view of the high frequency of outbursts which have been reported, and the excellent observational coverage, CG Dra was dropped from the ROP at the end of 2005 October. The ROP is intended for stars having outburst periods of more than a year and whose outburst characteristics are not well understood. However, CG Dra remains an intriguing star and fully warrants further attention from both visual and CCD observers.

## Acknowledgments

The authors would like to thank Dr Arne Henden, Director of the AAVSO, for permission to use data from the AAVSO International Database. In this regard, we gratefully acknowledge the contribution of all observers who have submitted

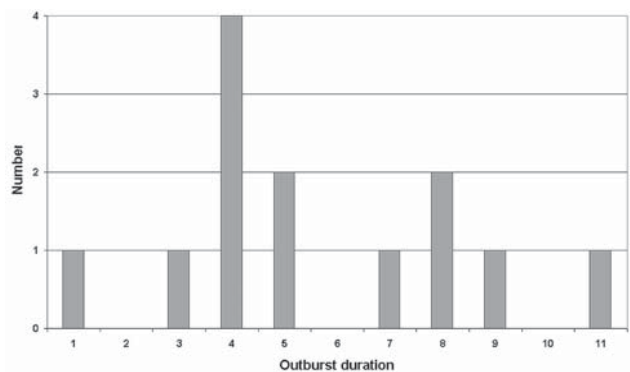

**Figure 4.** Histogram of the outburst duration (days).

**Table 2. Outbursts during the period JD 2453456 to 2453700 derived from VSS and AAVSO data**

| Outburst no. | Start of outburst, JD (+245000) | Duration, days | Time between outbursts, days |
|---|---|---|---|
| 1  | 3470.507 | 9  | 35.0 |
| 2  | 3505.465 | 3  | 13.1 |
| 3  | 3518.519 | 7  | 29.2 |
| 4  | 3547.740 | 5  | 10.8 |
| 5  | 3558.506 | 8  | 21.2 |
| 6  | 3579.682 | 1  | 9.7 |
| 7  | 3589.379 | 4  | 9.3 |
| 8  | 3598.660 | U  | 6.0 |
| 9  | 3604.629 | U  | 5.0 |
| 10 | 3609.615 | 8  | 16.7 |
| 11 | 3626.342 | 4  | 11.0 |
| 12 | 3637.375 | 4  | 11.1 |
| 13 | 3648.519 | 4  | 10.9 |
| 14 | 3659.420 | 5  | 10.9 |
| 15 | 3670.356 | 11 | |
|    |          | Mean 5.6 days | Median 11 days |

(U = undetermined)

data to the AAVSO International Database and BAA VSS. We would also like to thank Dr David Boyd for helpful comments about the statistical analysis of the data, and the referees, whose input has improved the quality of the paper.

**Addresses: JS:** 'Pemberton', School Lane, Bunbury, Tarporley, Cheshire, CW6 9NR, UK [bunburyobservatory@hotmail.com]
**RP:** 3 The Birches, Shobdon, Leominster, Herefordshire, HR6 9NG, UK [rdp@astronomy.freeserve.co.uk]
**GP:** 67 Ellerton Road, Kingstanding, Birmingham, B44 0QE, UK [Garypoyner@blueyonder.co.uk]